\documentclass{article}
\usepackage{amssymb}
\usepackage{amsmath}

\input{tcilatex}

\begin{document}

\title{Physics on the adiabatically changed Finslerian manifold and cosmology%
}
\author{ Anton A. Lipovka \\
Centro Investigacion en Fisica, Universidad de Sonora, \\
Sonora, Mexico. e-mail: aal@cifus.uson.mx }
\date{August 04, 2016 }
\maketitle

\begin{abstract}
In present paper we confirm our previous result [5] that the Planck constant
is adiabatic invariant of electromagnetic field propagating on the
adiabatically changed Finslerian manifold. Direct calculation from
cosmological parameters gives value h=6x10(-27) (erg s). We also confirm
that Planck constant (and hence other fundamental constants which depend on
h) is varied on time due to changing of geometry.

As an example the variation of the fine structure constant is calculated.
Its relative variation ((da/dt)/a) consist 1.0x10(-18) (1/s).

We show that on the Finsler manifold characterized by adiabatically changed
geometry, classical free electromagnetic field is quantized geometrically,
from the properties of the manifold in such manner that adiabatic invariant
of field is ET=6x10(-27)=h.

Equations of electrodynamics on the Finslerian manifold are suggested. It is
stressed that quantization naturally appears from these equations and is
provoked by adiabatically changed geometry of manifold. We consider in
details two direct consequences of the equations: i) cosmological redshift
of photons and ii) effects of Aharonov -- Bohm that immediately follow from
obtained equations. It is shown that quantization of system consists of
electromagnetic field and baryonic components (like atoms) is obvious and
has clear explanation.
\end{abstract}




\textbf{Cite this paper:}

\textit{Lipovka, A. (2017) Physics on the Adiabatically Changed Finslerian
Manifold and Cosmology. Journal of Applied Mathematics and Physics, 5,
582-595. doi: 10.4236/jamp.2017.53050.}

\section{Introduction}

The problem of formulation of general theory which could naturally unify
General Relativity (GR)\ and Quantum Theory (QT) is of most fundamental and
actual problem of modern physics. But to resolve this problem we should
first to know what is the Planck constant. This is the question which open
the door and allow us to find unified theory for all branches of physics. To
obtain this key we have first to have in mind some important facts.

1) In quantum mechanics, Planck constant always appears together with
derivatives in the same power. This fact clearly points out on the possible
relation between Planck constant and geometry.

2) Einstein [1] and later Debye [2] have shown from thermodynamical approach
that electromagnetic field is quantized alone, without any assumption on the
nature of oscillators. So this is another hint that it should be quantized
from geometry, without axiomatic introduction of the wave function.

3) Recently the cornerstone result was announced [3] in respect to the
quantization at very small scales. It was found that at the small scales a
quantum system behaves as a classical one (see paper [3] for details). This
result also argues that QT is pure geometrical phenomenon and it disappears
at small scales when geometry becomes to be approximately Euclidean.

4) In 2011 Zhotikov showed [4] that quantum postulates of Bohr and
quantization rules of Bohr-Sommerfeld, follow directly from the geometric
structure of space - time.

5) Recently we have published the paper in which we clearly points out the
geometrical origin of the Planck constant [5]. In this paper we have shown
that the Planck constant is adiabatic invariant of electromagnetic field on
the adiabatically changed Finslerian manifold. From this fact the variation
of the Planck constant on the time directly follows (and hence variations of
fundamental constants, for example variation of the fine structure constant,
due to adiabatically changed geometry [5,6]).

On the one hand we have these serious arguments for the geometric origin of
quantization, but on the other hand we have also serious problems with
Riemannian geometry. General relativity was created as theory on (pseudo-)
Riemannian manifold (we will use farther in this paper "Riemannian" instead
of "pseudo-Remannian"). Such choice was not favored by some serious
arguments but only by the fact the real geometry is curved. Metric tensor $%
g^{\mu \nu }(x)$ in this case depends only on the coordinates and this fact
applies some restrictions on the theory, and leads to the serious problem
with singularity. However, as it follows from experiments on measurement of
the cosmological constant, our Universe expands with acceleration and for
this reason complete geometry of Finsler should be favoured. On the
Finslerian manifold the metric tensor $g^{\mu \nu }(x,\overset{\cdot }{x})$
depends not only on coordinates, but also on velocities and (as it will be
shown in this paper) this fact allows us to introduce the cosmological
constant in natural way and to calculate from geometry correct value of the
Planck constant, unify gravity, electrodynamics and QT. The Hubble constant,
cosmological constant, quantization and much more follow from the Finsler
geometry in a natural way and we can conclude that Finsler geometry
naturally complies with all observational data.

In this paper we also obtain classical equations of motion for a system on
the manyfold with adiabatically changed geometry and supplied by Finslerian
metrical function $ds^{2}=g_{\mu \nu }(x,\overset{\cdot }{x})dx^{\mu
}dx^{\nu }$. We show how the Planck's constant naturally appears from
geometry and, within the (3+1) formalism, we write an exact relation between
the Planck constant on the one hand and scalar curvature and cosmological
constant on the other hand

We show how the classical electromagnetic field is naturally quantized due
to existence of adiabatic invariant of the field on adiabatically changed
Finslerian manifold.

Finally we write equations of electrodynamics, which classically
(geometrically) describe quantization of electromagnetic field (and hence of
any electromagnetically interacting system). To illustrate how it works, we
clearly considered two important particular cases which describe these
equations: i) case of free electromagnetic field, when quantization appears
from geometry and leads to losses of energy as photon propagates (so called
cosmological redshift), and ii) the Aharonov - Bohm effects, that
immediately follow from obtained electrodynamic equations.

All-around in this paper we suppose that Latin indexes $i,j,k,l,m=1,2,3$ and
greek $\alpha ,\beta ,...\mu ,\nu ...=0,1,2,3$ . Signature of metric is $%
(1,-1,-1,-1)$ .

\section{Adiabatic invariant and general formalism}

In classical physics the equations of motion follow immediately from the
relation $S_{m}(x)=S_{0}$ , where $S_{m}(x)$ is action of matter and $S_{0}$
characterize the 1-parameter family of hypersurfaces on M. By varying this
equation we obtain usual classical Hamilton (or Lagrange - Euler) equations
[7].

In the case of General Relativity we put into the right hand part of eq.(1)
the only invariant we have in pseudo-Riemannian geometry - scalar curvature $%
\mathcal{R}$ of the manifold and this way obtain the Einstein equations.

More general case appears when we consider a manifold supplied by Finslerian
metric. The case of static Finslerian manifold is well studied and can be
found for example in textbooks [8,9] and papers [10,11]. We should also
mention here very interesting discussion regarding unusual properties of
equations of motion on Finsler manifold suggested in [12]. Let us consider
what happens in general case, when the right-hand term of eq.(1) contains
some adiabatically changed parameters which characterize the Finslerian
manifold, i.e. when the geometry of the manifold is changed adiabatically on
time.

let $M$ be an $3$-dimensional, class $C^{3}$ manifold characterized by
scalar curvature $\mathcal{R}$ in the point $x^{k}$, where $x^{k}$ is a
local coordinate on an open subset $U\subset M$. Let also suppose $M$ be
supplied by Finsler metric $L(x,\overset{\cdot }{x})$ and write a
1-parameter family of hypersurfaces on the $M$ defined by equation

\begin{equation}
S_{m}(x)=S_{M}(x)  \tag{1}
\end{equation}

Here $S_{m}$\ stay for matter action and $S_{M}$\ corresponds to the 1 -
parameter\ family of the adiabatically changed hypersurfaces on the $M$. Our
more general case differs from that examined in [8] by the fact that now
hypersurface is changed adiabatically.

Our aim is to write the geodesic equations (Hamilton or Lagrange - Euler
equations) for this general case. By varying (1)

\begin{equation}
\delta S_{m}=\underset{}{\int \delta \mathcal{L}_{m}(x^{k},\overset{\cdot }{x%
}^{k})dt=}\int \delta (p_{k}dx^{k}-Hdt)=\delta S_{M}=\int \delta \mathcal{L}%
_{M}(x^{k},\overset{\cdot }{x}^{k})dt  \tag{2}
\end{equation}

we immediately obtain Hamilton - like equations

\begin{equation}
\frac{dp_{k}}{dt}=-\frac{\partial H}{\partial x^{k}}-\frac{\partial \mathcal{%
L}_{M}}{\partial x^{k}}  \tag{3}
\end{equation}

and

\begin{equation}
\frac{dx^{k}}{dt}=\frac{\partial H}{\partial p_{k}}+\frac{\partial \mathcal{L%
}_{M}}{\partial p_{k}}\text{ \ \ .}  \tag{4}
\end{equation}

One can see that due to expansion of the Universe (adiabatic changes of
right hand term in (1)) there appears an additional force in the equation
(3) and an additional velocity in (4), which we naturally can attribute to
cosmological constant (acceleration) and to the Hubble constant ($v=Hx$). It
actually corresponds to the fact that absolutely closed systems do not exist
and those additional terms appear due to adiabatic changes of geometry
(changes of metric tensor) which take place because of the Universe
expansion.

Absolutely the same way we can write equations of Lagrange - Euler by
varying (1):

\begin{equation}
\frac{\partial \mathcal{L}_{m}}{\partial x^{k}}-\frac{d}{dt}\frac{\partial 
\mathcal{L}_{m}}{\partial \overset{\cdot }{x}^{k}}=\frac{\partial \mathcal{L}%
_{M}}{\partial x^{k}}-\frac{d}{dt}\frac{\partial \mathcal{L}_{M}}{\partial 
\overset{\cdot }{x}^{k}}\text{ \ \ .}  \tag{5}
\end{equation}

As one can see in right part of this equation again appear two additional
terms due to expansion of the Universe (due to changing of geometry of the
manifold, as the system under consideration is moving).

\section{Exact Planck's constant value and quantization of electromagnetic
field}

Let us calculate value of the Planck constant from the parameters which
characterize the Finslerian manifold. Consider a generalized system
distributed over volume. Let $T_{p}(M)$ and $T_{p}^{\ast }(M)$ be
respectively tangent and cotangent bundles on $M$, where $p_{\alpha }\in
T_{p}(M)$ and $p^{\alpha }\in T_{p}^{\ast }(M)$ are covariant and
contravariant components of corresponding momentum.

We are interested here in the variation of the photon momentum due to
adiabatic change of geometry of $M$. This variation can be obtained directly
from the geometry [6], but also from Einstein's equations. Variation of the
momentum density of our system due to changes of M, in unit volume summed
over all directions is given by expression

\begin{equation}
\delta p=\frac{c^{3}}{8\pi G}\delta \mathcal{R}=\frac{c^{3}}{4\pi G}\delta 
\frac{1}{R^{2}}\text{ \ \ .}  \tag{6}
\end{equation}

Here, the scalar curvature $\mathcal{R=}2/R^{2}$\ and $R$\ is the radius of
curvature (we note here that we are interested in the evolution of the
electromagnetic field, which is described by a tensor of rank = 2. Such
tensors are known to correspond to a 2D surface). In this section, and later
in this article we are interested in the electromagnetic field, or more
precisely - its variation due to the change of the metric. Therefore, we
identify the coordinate $x$ with the size of the field resonator (a
\textquotedblleft box\textquotedblright ). Accordingly, the value $\overset{%
\cdot }{x}$ will characterize the speed of change for this \textquotedblleft
box\textquotedblright\ due to adiabatic changes in geometry of the manifold
M.

It is well known that on the Finsler manifold the metric tensor depends on $%
x $ and $\overset{\cdot }{x}$ for this reason the momentum $p^{2}=g^{\mu \nu
}(x,\overset{\cdot }{x})p_{\mu }p_{\nu }$ also depends on $x$ and $\overset{%
\cdot }{x}$. Let us assume that the geometry changes are adiabatic and due
to this fact we can represent $x$ as $x=x_{0}+\theta t$, where $x_{0}$ and $%
\theta $ are constant. In this case we have $\delta p(t,\overset{\cdot }{x})$
and, hence (see eq. (6)) $\delta R(t,\overset{\cdot }{x})$\ too. So we can
write

\begin{equation}
\delta p=\frac{-c^{3}}{2\pi G}\frac{1}{R^{3}}\delta R(t,\overset{\cdot }{x})=%
\frac{-c^{3}}{2\pi G}\frac{1}{R^{3}}\left( \delta R(t)+\frac{\partial R(H)}{%
\partial H}\delta H\right) \text{ \ \ ,}  \tag{7}
\end{equation}

here we note that $R(t,\overset{\cdot }{x})=R(t,Hx_{0})$ where $H$\ is the
Hubble constant. (We note here that we cannot measure the components of the
torsion tensor and the curvature of space, therefore, to calculate Planck's
constant and compare the obtained result with experiment, it is necessary to
re-express these quantities in terms of observables - the local values of
the Hubble constant and the cosmological constant. The indicated
calculations in this case will be valid only in the vicinity of our galaxy
cluster, therefore, to calculate Planck's constant in the general case, one
should take either the postulated curvature and torsion, or the values \$H\$
and \$\TEXTsymbol{\backslash}Lambda\$ observed in a particular case).

But by taking into account that (see [6])

\begin{equation}
\frac{\partial R}{\partial t}=\frac{\partial }{\partial t}\frac{c}{2H}=-%
\frac{c}{2H^{2}}\frac{\partial H}{\partial t}  \tag{8}
\end{equation}

one can write

\begin{equation}
\frac{\partial R}{\partial H}=\frac{\partial R}{\partial t}\frac{\partial t}{%
\partial H}=-\frac{c}{2H^{2}}\text{ \ \ .}  \tag{9}
\end{equation}

Let us consider variation of Hubble constant.

From the relation $\overset{\cdot }{x}=Hx$ we can find

\begin{equation}
\delta \overset{\cdot }{x}=x\delta H+H\delta x  \tag{10}
\end{equation}

or

\begin{equation}
x\delta H=\overset{\cdot \cdot }{x}\delta t-H\delta x  \tag{11}
\end{equation}

But the only cosmological acceleration we have experimentally measured is
that, associated with the cosmological constant $\Lambda $ , so we can write
for this variation

\begin{equation}
\delta H=\left( c^{2}\Lambda -H^{2}\right) \delta t\text{ \ \ .}  \tag{12}
\end{equation}

Substituting these expressions into (7) we find

\begin{equation}
\delta p=\frac{2c^{3}H}{\pi G}\left( \frac{2H^{2}}{c^{2}}-\Lambda \right)
\delta t\text{ \ \ ,}  \tag{13}
\end{equation}

or taking into account that for electromagnetic field $\mathcal{R=}2/R^{2}$
and by changing volume from the spherical coordinates to the euclidean ones,
we obtain

\begin{equation}
\delta p=\frac{3c^{3}H}{8\pi ^{2}G}\left( \mathcal{R}-4\Lambda \right)
\delta t\text{ \ \ .}  \tag{14}
\end{equation}

This is a variation of momentum (in unit volume in 3 directions) due to
adiabatically changed geometry, written for our generalized system localized
on the Finslerian manifold.

Now we are ready to write a complete adiabatic invariant for a free
propagating electromagnetic field. The components of the 4 - momentum $p$ of
free electromagnetic field propagating on the Finslerian manifold with
adiabatically changed geometry are varied on time. This variation proceeds
adiabatically and can be considered as linear function i.e. for energy $%
\varepsilon $ of the field, for example, we have

\begin{equation}
\frac{\delta \varepsilon }{\varepsilon }=-\frac{\delta t}{t}  \tag{15}
\end{equation}

so, the adiabatic invariant we are interested in is given by the expression

\begin{equation}
\varepsilon t=-\frac{\delta \varepsilon }{\delta t}t^{2}\text{ \ \ .} 
\tag{16}
\end{equation}

But

\begin{equation}
\delta \varepsilon =c\delta p_{k}\text{ \ \ .}  \tag{17}
\end{equation}

By substituting (14) into (16) we can write finally (we divide $\delta
p_{{}} $ by factor 3 because in the case of the electromagnetic plane wave,
we are interesting only in one direction of the momentum)

\begin{equation}
\varepsilon t=-ct^{2}\frac{\partial p}{\partial t}=-\frac{c^{4}H}{8\pi ^{2}G}%
\left( \mathcal{R}-4\Lambda \right) t^{2}=\eta ^{0}  \tag{18}
\end{equation}

so, by taking into account $\mathcal{R=}2/R^{2}$, $R=c/2H$\ (see comments
made before and also [6]), measured values of $H=73$ $kms^{-1}Mpc^{-1}=2.4%
\cdot 10^{-18}s^{-1}$ and $\Lambda =1.7\cdot 10^{-56}cm^{-2}$ [13] we have
for this adiabatic invariant $\eta ^{0}=h=6\cdot 10^{-27}\left( erg\cdot
s.\right) $ for one second and in cubic centimeter, as it should be. In the
same way we can obtain similar relations for other components of 4-momentum:

\begin{equation}
p_{\gamma }x^{\gamma }=\eta ^{\gamma }  \tag{19}
\end{equation}

(there is no summation over $\gamma $\ in this relation and for the photon
propagating in direction $x^{3}$ the components $p_{1}=p_{2}=0$). In the
case of plane electromagnetic wave, introduced here 4-vector $\eta ^{\gamma
} $\ has components in unit volume:

\begin{equation*}
\eta ^{\gamma }=(h,0,0,h)\text{ \ \ .}
\end{equation*}

\bigskip And for general case we can write

\begin{equation}
\eta ^{\gamma }=(h,h,h,h)\text{ \ \ .}  \tag{20}
\end{equation}

Here the adiabatic invariant for electromagnetic field (Planck constant),
which depends clearly on the parameters of the manifold $\mathcal{R}$ and $%
\Lambda $\ (and consequently depends on time) is:

\begin{equation}
h=-\frac{c^{4}H}{8\pi ^{2}G}\left( \mathcal{R}-4\Lambda \right) t_{0}^{2} 
\tag{21}
\end{equation}

where $H$ is the Hubble constant, $\mathcal{R}$ is the scalar curvature
discussed above (see discussion after expression (6)) $\Lambda $ is the
cosmological constant and the constant $t_{0}$ is equal to one second.
Expression (21) gives for unit time and unit volume at present epoch $%
h=6\cdot 10^{-27}\left( erg\cdot s.\right) $ as it was mentioned above. From
this relation it is easy to see that the Planck constant depends on time as $%
h\sim 1/t$.

\subsection{Hilbert integral}

Now let's consider integral of Hilbert for particular case of the free
electromagnetic field propagating along geodesic on the adiabatically
changed Finslerian manifold:

\begin{equation}
\underset{}{\int }p_{\alpha }dq^{\alpha }=\Delta S_{M}\text{ \ \ ,}  \tag{22}
\end{equation}

where $p_{\alpha }$\ is 4-momentum of the field, $q^{\alpha }$\ is
generalized coordinate and right hand term $\delta S_{M}$, as before,
corresponds to the changing of the 1-parameter family of hypersurfaces due
to adiabatic variation of geometry as system under consideration is moving
on $M$ . For electromagnetic field with Lagrangian $\mathcal{L}_{m}=F_{\mu
\nu }F^{\mu \nu }/16\pi $\ we have from (22)

\begin{equation}
\underset{}{\frac{1}{c}\int }\frac{\partial \mathcal{L}_{m}}{\partial A_{\mu
,\nu }}A_{\mu ,\sigma }dx^{\sigma }=\Delta S_{M}  \tag{23}
\end{equation}

By taking into account that tensor of energy - momentum is

\begin{equation}
T_{\sigma }^{\nu }=\frac{\partial \mathcal{L}_{m}}{\partial A_{\mu ,\nu }}%
A_{\mu ,\sigma }-\delta _{\sigma }^{\nu }\mathcal{L}_{m}  \tag{24}
\end{equation}

we can write for propagating classical electromagnetic field

\begin{equation}
\frac{1}{c}\int \left( T_{\sigma }^{\nu }+\delta _{\sigma }^{\nu }\mathcal{L}%
_{m}\right) dx^{\sigma }=\Delta S_{M}  \tag{25}
\end{equation}

If the field propagates in the direction $x_{3}$, than electric field $%
\mathbf{E}=E_{1}$ and magnetic field $\mathbf{H}=H_{2}$. So the only
non-zero components of the field tensor $F_{\mu \nu }$ are $%
F_{01}=-F_{10}=E_{1}$ ; $F_{13}=-F_{31}=-H_{2}$ and hence $%
T^{00}=T^{33}=T^{30}=(E^{2}+H^{2})/8\pi $ , where for the plane wave we have 
$E=E_{0}cos(k_{\alpha }x^{\alpha })$ and $H=H_{0}sin(k_{\alpha }x^{\alpha })$
.

In this case for the $00$ - component, for example, by taking into account
(18) and after elementary integration we obtain

\begin{equation}
\frac{E_{0}^{2}+H_{0}^{2}}{8\pi }=\frac{h}{T}=h\nu \text{ \ \ .}  \tag{26}
\end{equation}

Similar relations one can write for other components.

So, as one can see the classical electromagnetic field is quantized due to
adiabatic variation of the Finslerian manifold and we do not need some
artificial methods to quantize it.

\subsection{Variation of the fine structure constant}

As it was shown before [6] even on the adiabatically changed Riemannian
manifold the value of the fine structure constant is changed adiabatically
on time. This variation appears due to a change in the metric of space, in
which the atom is localized. In the case of Finslerian manifold this
variation is smaller by factor 2/3 due to the presence of the cosmological
constant. To show this let's start from (14). For one direction (divided by
factor 3) we have from (14):

\begin{equation}
\delta p=\frac{c^{3}H}{8\pi ^{2}G}\left( \mathcal{R}-4\Lambda \right) \delta
t\text{ \ \ .}  \tag{27}
\end{equation}

But the fine structure constant is $\alpha =V/c$\ where $V$\ is electron
velocity at the first Bohr orbit. Momentum in this case we can write as

\begin{equation}
P=\frac{m\alpha c}{\sqrt{1-\alpha ^{2}}}  \tag{28}
\end{equation}

so

\begin{equation}
\delta P=\frac{mc}{\left( 1-\alpha ^{2}\right) ^{3/2}}\delta \alpha  \tag{29}
\end{equation}

and

\begin{equation}
\delta \alpha =\frac{\left( 1-\alpha ^{2}\right) ^{3/2}Hc^{3}}{mc8\pi ^{2}G}%
\left( \mathcal{R}-4\Lambda \right) \delta t  \tag{30}
\end{equation}

that give us value $\overset{\cdot }{\alpha }/\alpha =-1.03\cdot 10^{-18}$
(for 1 second), $\mathcal{R=}2/R^{2}$, $R=c/2H$\ (see [6]), $H=73$ $%
kms^{-1}Mpc^{-1}=2.4\cdot 10^{-18}s^{-1}$ and $\Lambda =1.7\cdot
10^{-56}cm^{-2}$).

\section{\protect\bigskip Electrodynamics on the Finslerian manifold}

In Riemannian geometry, the first pair of equations of electrodynamics
follows directly from the properties of the field tensor.

\begin{equation}
F_{\mu \nu }=A_{\nu ;\mu }-A_{\mu ;\nu }=A_{\nu ,\mu }-A_{\mu ,\nu } 
\tag{31}
\end{equation}

where

\begin{equation}
A_{\mu ;\nu }=A_{\mu ,\nu }-\Gamma _{\mu \nu }^{\sigma }A_{\sigma }  \tag{32}
\end{equation}

And for this reason ($\Gamma _{\mu \nu }^{\sigma }=\Gamma _{\nu \mu
}^{\sigma }$) the first pair of equations on Riemannian manifold with
constant scalar curvature is

\begin{equation}
\partial _{\sigma }F_{\mu \nu }+\partial _{\mu }F_{\nu \sigma }+\partial
_{\nu }F_{\sigma \mu }=0  \tag{33}
\end{equation}

On \ the Finsler manifold, we can obtain the first pair in the same way, but
in this case the field tensor is

\begin{equation}
\overset{\backsim }{F}_{\mu \nu }=A_{\nu ;\mu }-A_{\mu ;\nu }  \tag{34}
\end{equation}

where covariant differentials $DA_{\mu }$\ include now terms with the Cartan
connections $C_{\mu \nu \sigma }^{{}}=\frac{1}{2}\frac{\partial g_{\mu \nu }%
}{\partial \overset{\cdot }{x}^{\sigma }}$ like this $C_{\mu \text{ }\nu }^{%
\text{ }\sigma }A_{\sigma }d\overset{\cdot }{x}^{\nu }$\ and also $\Gamma
_{\mu \text{ }\nu }^{\text{ }\sigma }A_{\sigma }dx^{\nu }$. In the most
important case, we are interested in, when the scalar curvature is small and
the metric tensor has spatial structure described by the Robertson-Walker
metric, the additional terms $C_{\mu \nu }^{\sigma }A_{\sigma }d\overset{%
\cdot }{x}^{\nu }$\ and $\Gamma _{\mu \nu }^{\sigma }A_{\sigma }dx^{\nu }$\
in covariant differentials can be evaluated easily as $\approx A_{\sigma
}dx^{\nu }/R$\ \ (here $1/R=2H/c=5\cdot 10^{-29}$ is the inverse radius of
scalar curvature at the point of observation)\ so we have

\begin{equation}
\overset{\backsim }{F}_{\mu \nu }=A_{\nu ,\mu }-A_{\mu ,\nu }-t_{\mu \nu } 
\tag{35}
\end{equation}

where our estimation for the small components consists $dt_{\mu \nu }\approx
A_{\mu }dx^{\nu }/R\approx A_{\mu }dx^{\nu }\cdot 10^{-29}$.

As one can see these components probably will be significant only in the
vicinity of black holes and can be omitted in our present consideration.

For this reason the first pair of the equations can be written as

\begin{equation}
\overset{\sim }{F}_{\mu \nu ;\sigma }+\overset{\sim }{F}_{\nu \sigma ;\mu }+%
\overset{\sim }{F}_{\sigma \mu ;\nu }=0  \tag{36}
\end{equation}

or, by taking into account our estimations discussed above, we can write

\begin{equation}
\partial _{\sigma }F_{\mu \nu }+\partial _{\mu }F_{\nu \sigma }+\partial
_{\nu }F_{\sigma \mu }=O(t_{\mu \nu ,\sigma })\text{ \ \ .}  \tag{37}
\end{equation}

So one can see that the first pair of electrodynamic equations remains to be
the first pair of the Maxwell equations with high precision.

The second pair of equations of electrodynamics follows directly from
variation of functional (1) if we consider a charge characterized by
4-current $j^{\alpha }$, and the electromagnetic field on the Finslerian
manifold:

\begin{equation}
S_{m}=S_{M}  \tag{38}
\end{equation}

Here $S_{M}$\ as before, corresponds to the family of the hypersurfaces on
the expanded manifold. By varying $S_{m}$ we have

\begin{equation}
\delta S_{m}=-\frac{1}{c}\underset{\Omega }{\int }\left[ \frac{1}{c}%
j^{\alpha }\delta A_{\alpha }+\frac{1}{16\pi }\delta (F_{\mu \nu }F^{\mu \nu
})\right] d\Omega  \tag{39}
\end{equation}

here we put $F_{\mu \nu }$ instead of $\overset{\backsim }{F}_{\mu \nu }$\
because, as we have seen, small additional terms corresponding to small
components of $t_{\mu \nu }$ are insignificant in the case of
Robertson-Walker metric. Integrating the second term by parts, we obtain

\begin{equation}
\delta S_{m}=-\frac{1}{c}\underset{\Omega }{\int }\left[ \frac{1}{c}j^{\mu }+%
\frac{1}{4\pi }\frac{\partial F^{\mu \nu }}{\partial x^{\nu }}\right] A_{\mu
,\sigma }\delta x^{\sigma }d\Omega  \tag{40}
\end{equation}

By varying $S_{M}$ we have (see eq. (16 - 20))

\begin{equation}
\delta S_{M}=\underset{\Omega }{\int }\frac{\eta ^{\sigma }}{(x^{\sigma
})^{2}}\delta x^{\sigma }d\Omega  \tag{41}
\end{equation}

where, as it was shown before, $\eta ^{\sigma }=(h,h,h,h)$ \ in unit volume
(here $h$ is the Planck constant) and $\Omega $\ is a 4-volume. The
equations under discussion one can write as follows

\begin{equation}
\frac{1}{c}\underset{\Omega }{\int }\left[ \frac{1}{c}j^{\mu }+\frac{1}{4\pi 
}\frac{\partial F^{\mu \nu }}{\partial x^{\nu }}\right] A_{\mu ,\sigma
}\delta x^{\sigma }d\Omega =\underset{\Omega }{-\int }\frac{\eta ^{\sigma }}{%
(x^{\sigma })^{2}}\delta x^{\sigma }d\Omega +O\left( (\eta ^{\sigma
})^{2}\right)  \tag{42}
\end{equation}

or finally

\begin{equation}
\frac{1}{c}\left[ \frac{1}{c}j^{\mu }+\frac{1}{4\pi }\frac{\partial F^{\mu
\nu }}{\partial x^{\nu }}\right] A_{\mu ,\sigma }=-\frac{\eta ^{\sigma }}{%
(x^{\sigma })^{2}}+O((\eta ^{\sigma })^{2})  \tag{43}
\end{equation}

(there is no summation over $\sigma $ here).

This is the second pair of equations of electrodynamics on the adiabatically
changed Finslerian manifold. The bounded electromagnetic field (second term)
in this case is explicitly included into consideration, as it takes place in
the case of Bohmian formalism when this field appears in QT as quantum
potential (see [14] for details and also results of paper [15]). It is easy
to see that when the Planck constant tends to zero, the expression (43) (as
it should be) is converted into the Maxwell equations. (Especially emphasize
here that in this extreme case, the topology becomes simple, space is
isotropic, flat and described by the Minkowski geometry). Let us consider
two important cases which immediately follow from these equations.

\subsection{Cosmological redshift}

It is well known that as the photon propagates through expanding universe
its frequency (or wave length) is changed. This loss of energy by free
electromagnetic field, named as cosmological redshift, appears in our
equations by natural way as losses of the energy by photon due to
adiabatically changed geometry of manifold.

\begin{equation}
\frac{1}{4\pi c}\frac{\partial F^{\mu \nu }}{\partial x^{\nu }}A_{\mu
,\sigma }=\frac{1}{8\pi c}\frac{\partial (F_{\mu \nu }F^{\mu \nu })}{%
\partial x^{\sigma }}=-\frac{\eta ^{\sigma }}{(x^{\sigma })^{2}}+O((\eta
^{\sigma })^{2})  \tag{44}
\end{equation}

(there is no summation over $\sigma $).

The meaning of these equations is most obvious if we put $\sigma =0$. In
this case, the left side will be the electromagnetic field energy loss with
time, and the right side became $h/t^{2}$ - small value associated with the
geometry of the universe (see expression (18)), i.e.

\begin{equation}
\frac{1}{8\pi }\frac{\partial (F_{\mu \nu }F^{\mu \nu })}{\partial t}\approx
-\frac{h}{t^{2}}=\frac{c^{4}H}{8\pi ^{2}G}\left( \mathcal{R}-4\Lambda \right)
\tag{45}
\end{equation}

\subsection{The Aharonov - Bohm effects}

Another important case that follows directly from the second pair of
equations is the Aharonov - Bohm effects. As it is known, a necessary
condition for the existence of the Aharonov-Bohm (AB) effects is the
presence, in the overall structure of the equations, of the "zero field"
potentials which cannot be removed by gauge transformations and they do not
create electromagnetic fields [16,17]. These "zero - potentials" are the
result of "non-trivial topology" of the area on which the particle moves
[16,18,19]. Such a situation arises in electrodynamics of anisotropic media
where the structure of Maxwell's equations eliminates the possibility of
satisfying the boundary conditions. To satisfy regularly the boundary
conditions in anisotropic media, usually the zero - potential is introduced
, which do not create electromagnetic fields (see [16] and references
therein). In the case of adiabatically expanding Finslerian manifold, the
anisotropy of space occurs for any moving body automatically as right part
of eq. (43).Therefore, it is safe to say that in the case of AB effects we
are dealing directly with the anisotropy of space, which appears due to
adiabatically changed Finslerian manifold as the particle moves along its
trajectory. In this case, the role of the zero potentials (which do not
generate electromagnetic fields) performs variable geometry of space as it
follows directly from (43).

In absence of electric and magnetic fields en route of propagation of the
particle under consideration, the second term disappears (but it still take
place inside of the solenoid and affects our particle: "In spite of the fact
that the magnetic field vanishes out of the solenoid, the phase shift in the
wave functions is proportional to the corresponding magnetic flux inside of
the solenoid" [20] ) and we obtain (we neglect here by the small term $%
O((\eta ^{\sigma })^{2})$)

\begin{equation}
\frac{1}{c^{2}}j^{\mu }A_{\mu ,\sigma }=-\frac{\eta ^{\sigma }}{(x^{\sigma
})^{2}}  \tag{46}
\end{equation}

but $j^{\mu }=(\rho c,\rho V^{k})$\ (here $\rho $\ is charge density and $%
V^{k}$\ is 3-velocity) so if we put $\rho =e$\ and remember that $\delta
j^{\mu }=0$ (for this reason $A_{\mu }\partial _{\sigma }j^{\mu }=0$\ and $%
A_{\mu ,\sigma }j^{\mu }=\partial _{\sigma }(A_{\mu }j^{\mu })$) we obtain
by using the Gauss theorem

\begin{equation}
\frac{e}{c}A_{0}=-\frac{\eta ^{0}}{(x^{0})}\text{ \ \ , \ \ }\frac{e}{c}%
A_{k}V^{k}=-\frac{\eta ^{k}}{(x^{k})}\text{ \ \ .}  \tag{47,48}
\end{equation}

These equations describe the electric and magnetic effects of Aharonov -
Bohm (here $x^{0}$\ and $x^{k}$\ are fixed). Namely for $\mu =0$\ we have
for the phase variation $\Delta \Phi $

\begin{equation}
e\underset{t}{\int }\varphi dt=-h\frac{\Delta t}{1\sec .}=-h\Delta \Phi 
\tag{49}
\end{equation}

electric effect of Aharonov - Bohm, and when $\mu =k$\ \ (here $k=1,2,3$) we
have relation

\begin{equation}
\frac{e}{c}\underset{l}{\int }A_{k}dx^{k}=-h\frac{\Delta x}{1cm.}=-h\Delta
\Phi  \tag{50}
\end{equation}

describes magnetic effect of Aharonov - Bohm.

To conclude this part we would like to stress again that whereas the bounded
field (second term in (43)) do not appears in these relations, it actually
affects the moving particle through potentials $A_{\mu }$\ [20] and this
bounded field corresponds to quantum potential in the Bohmian formalism
[14,15].

\section{Complete theory}

In previous part we have obtained electrodynamic equations. They are applied
in the case when the movement of charge (or 4-current $j^{\mu }$ ) is
defined. To construct self-consistent theory we should treat $j^{\mu }$ as
variable\ from the beginning.

We consider now a charge characterized by 4-current $j^{\alpha }$, and the
electromagnetic field on the Finslerian manifold:

\begin{equation}
S_{m}=S_{M}  \tag{51}
\end{equation}

Here $S_{M}$\ as before corresponds to family of the hypersurfaces on the
expanded manifold. In complete form, when $\delta j^{\mu }\neq 0$ , we have
the action

\begin{equation}
-\sum \int mcds-\frac{1}{c}\underset{\Omega }{\int }\left[ \frac{1}{c}%
j^{\alpha }A_{\alpha }+\frac{1}{16\pi }(F_{\mu \nu }F^{\mu \nu })\right]
d\Omega =S_{M}  \tag{52}
\end{equation}

which describes quantum properties of our system. It is clear there are a
lot of different systems, particular cases and applications which can not be
considered here because of their huge amount. For this reason let us
consider here the hydrogen atom as an example. In order to coincide with the
Schr\"{o}dinger formulation of quantum mechanics, we should neglect the
third term in (52) which corresponds to quantum potential [14,15] in Bohmian
formulation, and gives the zero energy correction (for example in the case
of harmonic oscillator it gives term 1/2 in expression for energy [5,14]).
In this case by varying (52) we have

\begin{equation}
-\delta \int (mcds+\frac{e}{c}A_{\alpha }dx^{\alpha })=\delta S_{M}  \tag{53}
\end{equation}

and we can write

\begin{equation}
\int \left[ mc\frac{du_{\mu }}{ds}-\frac{e}{c}\left( \frac{\partial A_{\nu }%
}{\partial x^{\mu }}-\frac{\partial A_{\mu }}{\partial x^{\nu }}\right)
u^{\nu }\right] \delta x^{\mu }ds=-\underset{}{\int }\frac{\eta ^{\mu }}{%
(x^{\mu })^{2}}\delta x^{\mu }ds\text{ \ \ .}  \tag{54}
\end{equation}

So, finally we obtain equations of motion

\begin{equation}
mc\frac{du_{\mu }}{ds}-\frac{e}{c}F^{\mu \nu }u_{\nu }=-\frac{\eta ^{\mu }}{%
(x^{\mu })^{2}}\text{ \ \ .}  \tag{55}
\end{equation}

By taking into account that classical period for orbital movement of
electron is

\begin{equation}
T=\pi e^{2}\sqrt{\frac{m}{2\left\vert E\right\vert ^{3}}}\text{ ,}  \tag{56}
\end{equation}

in classical limit $v<<c$ the straightforward calculations give the energy
for first Bohr orbit, obtained from classical electrodynamic on Finslerian
manifold: \ 

\begin{equation}
E_{1}=\frac{me^{4}}{2\hbar ^{2}}  \tag{57}
\end{equation}

that coincides with quantum calculations. Relativistic corrections are
obvious and follow from (55).

\section{Conclusions}

In this paper we confirm our previous result [5] that Planck constant is
adiabatic invariant of electromagnetic field propagating on the
adiabatically changed Finslerian manifold. Direct calculation of the Planck
constant value made from cosmological parameters gives $h=6\cdot
10^{-27}\left( ergs\cdot s.\right) $ that is in excellent agreement with the
measured value. We also confirm that Planck constant (and hence other
fundamental constants which depend on h) is varied on time due to changing
of geometry of the manifold.

As an example we suggest a calculation of the fine structure constant
variation. The obtained value consist $\overset{\cdot }{\alpha }/\alpha
=-1.03\cdot 10^{-18}$ (for 1 second) and this variation is expected to be
measured in nearest future.

We show that on the Finslerian manifold characterized by adiabatically
changed geometry, classical free electromagnetic field is quantized
geometrically, from the properties of the manifold.

Equations for electrodynamics on the Finslerian manifold are suggested. It
is shown that quantization naturally appears from these equations and is
resulted from adiabatically changed geometry of manifold. We consider in
details two direct consequences of the equations - cosmological redshift of
photons and effects of Aharonov - Bohm.

Finally we show appearance of quantization of systems consisting of
electromagnetic field and charged baryonic components (like atoms and
molecules).

\section{\protect\bigskip Acknowledgments}

I would like to express my deepest appreciation and gratitude to my
teachers: Choban E.A., Dubov V.V., Dubrovich V.K., Finkelstain B.A.,
Gosachinsky I.V., Ivanov V.K., Khersonsky V.K., Okunev A.A., Tichomirov
S.R., Toptygin I.N., Rukolaine A.V., Varshalovich D.A., Zlatin A.N., and
many others. I also have to thank my wife Olga. Thanks to her efforts, I was
able to complete this work for the summer vacation 2016.

\end{document}